\documentclass[preprint]{jpsj2}

\renewcommand{\theenumi}{( \( \alph{enumi} \) )}
\def\runtitle{A Recursive Method of the Stochastic State Selection 
for Quantum Spin Systems }
\def\runauthor{Tomo  \textsc{Munehisa} and Yasuko \textsc{Munehisa}}

\title{%
A Recursive Method of the Stochastic State Selection \\ 
for Quantum Spin Systems  }

\author{%
Tomo  \textsc{Munehisa} and
Yasuko \textsc{Munehisa}
}

\inst{%
 Faculty of Engineering, Univ. of Yamanashi, Kofu 400-8511
}
\recdate{\today}

\abst{%
In this paper we propose 
{\em the recursive stochastic state selection method},
an extension of the recently developed 
{\em stochastic state selection method} in Monte Carlo 
calculations for quantum spin systems.   
In this recursive method we use intermediate states to define
probability functions for stochastic state selections. 
Then we can diminish variances of samplings when we calculate  
expectation values of the powers of the Hamiltonian.
In order to show the improvement 
we perform numerical calculations of the spin-$1/2$ anti-ferromagnetic 
Heisenberg model on the triangular lattice.
Examining results on the ground state of the 21-site system we confide
this method in its effectiveness. 
We also calculate the lowest and the excited energy eigenvalues as well
as the static structure factor for the 36-site system.
The maximum number of basis states kept in a computer memory for this
system is about $3.6 \times 10^7$.
Employing a translationally invariant initial trial state, we evaluate 
the lowest energy eigenvalue within $0.5 \%$ of the statistical errors. 
}

\kword{%

quantum spin, large size, numerical calculation, Monte Carlo,
triangular lattice
}
\begin{document}
\sloppy
\maketitle

\section{Introduction}

Quite recently we have proposed a new method of Monte Carlo
calculations, which we call {\em the stochastic state selection} 
(SSS) method\cite{mune}, to obtain energy eigenvalues of quantum spin systems.
Under this method we can numerically eliminate a lot of basis states 
in a mathematically justified way.
This reduction enables us to calculate expectation values of
powers of the Hamiltonian even when a limited computer memory is available.
In a previous work\cite{mune2} we applied this method to the 
two-dimensional Shastry-Sutherland model and obtained reliable energy 
eigenvalues for the first excited states 
on a $8 \times 8$ lattice {\em across the critical region}.

In this paper we propose an extended version of the SSS method, 
which we will refer to as  
{\em the recursive stochastic state selection} (RSSS) method.
In the SSS method we have introduced 
{\em random choice matrices}, whose elements are 
stochastic variables to follow probability 
functions named {\em on-off probability functions}. Here these probability
functions are defined only once at the beginning. 
In contrast to them, 
the on-off probability functions in the RSSS method are defined 
in sequence for each random choice matrix. 
By this extension we can expect smaller variances of the expectation
values of high powers of the Hamiltonian compared to the
SSS results. Here we examine the effectiveness of the RSSS method 
in numerical studies of the anti-ferromagnetic 
Heisenberg quantum spin-$1/2$ system on the triangular 
lattice\cite{oguchi,nishim,leung,sindz,bernu,capri,farnell}.
To this typical strongly frustrated system, 
other methods such as the ordinary Monte Carlo 
methods or the perturbative calculations are hardly applicable.
We first show that the method is successful for the model on the lattice of 21
sites. Then we proceed to the 36-site system to demonstrate that 
we can estimate the ground state energy eigenvalue, 
a first-excited energy eigenvalue and 
the static structure factor by this method.

The plan of this paper is as follows.
In the next section we briefly summarize the SSS method in order to
prepare for the following section and the Appendix. 
Section 3 is to explain the RSSS method, where the recursive procedure of the
method is stated.
In sections 4 and 5 we apply the method to the
anti-ferromagnetic Heisenberg spin-$1/2$ system on the triangular
lattice. Section 4 is devoted to basic examinations.
We first examine results on the energy eigenvalue of the ground
state for a small system of 21 sites.
We show that the RSSS results are in good agreement with exact values. 
We also observe much less variances for the expectation values 
of high powers of the Hamiltonian compared to the SSS results.
Then we calculate the lowest energy eigenvalue for the  
36-site system. Here we discuss on a way to find a good approximate state.
In addition, we state how we estimate the ground state energy eigenvalue
from numerically calculated expectation values. 
In section 5 we present further calculations for the 36-site system. 
Here we employ an improved initial trial state which reflects the 
translational invariance of the Hamiltonian. 
We show that the statistical errors of the ground state 
energy are much reduced by this improvement. 
Our result is $E = -20.04 \pm \begin{array}{l} 
{\mbox{\small 0.10}} \\ {\mbox{\small 0.06}} \end{array}$, 
which is about $0.6 \%$ higher
than the known exact value $E = -20.17344$\cite{leung, bernu,capri}.
We also present results on the energy eigenvalue of an excited state
and the static structure factor. 
The final section is devoted to summary and discussions.
In the Appendix we discuss some properties concerning to 
the on-off probability functions.

\section{Stochastic State Selection Method}

This section is devoted to a brief summary of the SSS method\cite{mune}. 
Let $\mid \psi \rangle$ be a state of the system 
which is expanded by a basis \{ $\mid  i \rangle$\} as 
\begin{eqnarray}
\mid  \psi \rangle = \sum_{i=1}^{N_{\rm V}} \mid  i \rangle c_i \ ,
\label{expnd}
\end{eqnarray}
$N_{\rm V}$ being the size of the full vector space.
We denote the Hamiltonian of the system by $\hat H$.
In order to numerically evaluate the expectation values
\begin{eqnarray} 
E(L) \equiv \langle \psi \mid \hat H ^L \mid \psi \rangle  \ \ \ \ \ \
(L=1, 2, \cdots) \ , 
\end{eqnarray}
we define\cite{foot1} 
\begin{eqnarray}
  E_{\{\eta\}}(L) \equiv \langle \psi \mid \hat H
M_{\{ \eta ^{(L)}\}} \hat H M_{\{ \eta ^{(L-1)}\}} \cdots \hat H
M_{\{ \eta ^{(1)}\}} \mid  \psi \rangle \ .
\label{eldeff}
\end{eqnarray}
The random choice matrices 
$M_{\{ \eta ^{(m)}\}}$ $(m=1,2,\cdots,L)$ are the following diagonal matrices 
\begin{eqnarray}
M_{ \{ \eta^{(m)} \}} \equiv
\left( \begin{array}{cccc} \eta_1^{(m)} & 0 & \cdots & 0 \\
0 & \eta_2^{(m)} & \cdots & 0  \\
\cdots & \cdots & \cdots & \cdots \\
 0 & 0 & \cdots & \eta_{N_{\rm V}}^{(m)} \end{array}
\right) \ ,
\label{mmdef}
\end{eqnarray}
with random variables $\eta_i^{(m)}$'s. 
These variables are determined according to the on-off probability functions 
\begin{eqnarray}
P_i(\eta) \equiv P(\eta; a_i) \ , \ \ \ \ \ \
a_i \equiv \left\{ \begin{array}{ll} 
{\rm max} \left( 1, \epsilon/|c_i| \right)  & (c_i \neq 0 ) \\ 
\epsilon / \delta & (c_i = 0) \end{array} \right. \ ,
\label{picdef}
\end{eqnarray}
where 
\begin{eqnarray}
P(\eta; a) &\equiv&
\frac{1}{a}\delta \left(\eta -a \right) +
\left( 1- \frac{1}{a}\right) \ \delta \left( \eta \right) \ ,
\label{pdef}
\end{eqnarray}
and $\epsilon$ and $\delta ( < \epsilon )$ denote given positive constants.
Note that each random choice matrix is {\em independently} generated 
using the {\em same} on-off probability
function $ P_i(\eta) $ for $\eta_i^{(m)}$. We can reduce the
effective number of the basis by operating $M_{ \{ \eta^{(m)} \}}$
to the state $\hat H M_{\{ \eta ^{(m-1)}\}} \cdots \hat H
M_{\{ \eta ^{(1)}\}} \mid  \psi \rangle $ 
because many of generated values of $\eta_i ^{(m)}$'s in 
$\{ \eta ^{(m)} \} \equiv \{ \eta_1 ^{(m)}, \eta_2 ^{(m)}, \cdots,
\eta_{N_{\rm V}} ^{(m)} \}$ are zero.
It is guaranteed that the statistical average of 
$E_{\{\eta \}}(L)$ gives $E(L)$,
\begin{eqnarray}
\langle \! \langle E_{\{\eta \}}(L) \rangle \! \rangle =  E(L) \ ,
\label{eleel}
\end{eqnarray}
since the statistical average of $\eta_i^{(m)}$ is  
\begin{eqnarray}
\langle \! \langle   \eta_i^{(m)} \rangle \! \rangle 
&\equiv& 
\int_0^\infty \eta_i^{(m)}  P_i(\eta_i^{(m)}) d \eta_i^{(m)} = 1 \ , 
\label{etai1} 
\end{eqnarray} 
and 
$\langle \! \langle  \eta_i^{(1)}  \eta_j^{(2)}  \cdots   \eta_l^{(L)}
\rangle \! \rangle = \langle \! \langle  \eta_i^{(1)}
\rangle \! \rangle \langle \! \langle  \eta_j^{(2)}\rangle \! \rangle  \cdots
\langle \! \langle \eta_l^{(L)}\rangle \! \rangle$ 
holds for independently generated 
$\{ \eta^{(1)} \}$, $\{ \eta^{(2)} \}$, $\cdots$, $\{ \eta^{(L)} \}$.

\section{Recursive Stochastic State Selection Method}

In this section we show a procedure to estimate expectation values
following the RSSS method.
While the on-off probability functions (\ref{picdef}) in the SSS method 
are {\em common} to {\em all} $m$ for each $i$, 
we generate the random variables $\{ \eta^{(m)} \}$ in the RSSS method using 
the preceding intermediate state $\hat H M_{\{ \eta ^{(m-1)}\}} \cdots \hat H
M_{\{ \eta ^{(1)}\}} \mid  \psi \rangle$.

We start from the state 
$\mid \psi^{(0)} \rangle \equiv \mid \psi \rangle$, 
\begin{eqnarray}
\mid \psi^{(0)} \rangle = \sum_{i=1}^{N_{\rm V}} \mid  i \rangle c^{(0)}_i \ .
\label{expnd0}
\end{eqnarray}
Then we determine the {\em first} random choice matrix $M_{\{\eta ^{(1)}\}}$ 
generating random variables $\eta^{(1)}_i$ according to the  
on-off probability functions    
\begin{eqnarray}
P_i^{(1)} \left( \eta  \right) \equiv P \left( \eta; a ^{(1)}_i  \right) \ ,
\ \ \ \ \ \ a^{(1)}_i \equiv 
\left\{ \begin{array}{ll} 
{\rm max} \left( 1, \epsilon/|c^{(0)}_i| \right)  & (c^{(0)}_i \neq 0 ) \\ 
\infty & (c^{(0)}_i = 0) \end{array} \right. \ ,     
\label{pi1}
\end{eqnarray}
where the function $P(\eta ; a)$ is given by (\ref{pdef})   
and $\epsilon$ is a positive constant.  
Calculating the state 
$\hat{H} M_{\{\eta ^{(1)}\}} \mid \psi^{(0)}\rangle $ 
with this $M_{\{\eta ^{(1)}\}} $ we obtain  
$\mid \psi^{(1)} \rangle \equiv
 \hat{H} M_{\{\eta ^{(1)} \}} \mid \psi^{(0)}\rangle /C^{(1)}$.
Here $C^{(1)} (>0)$ is given by the equation 
\begin{eqnarray}
\left[ C^{(1)} \right] ^2 = \langle \psi^{(0)} \mid  M_{\{\eta ^{(1)} \}}
 \hat{H}^2 M_{\{\eta ^{(1)}\}} \mid \psi^{(0)} \rangle \ , 
\end{eqnarray}
which follows from the normalization condition 
$\langle  \psi^{(1)} \mid \psi^{(1)} \rangle = 1$. 
Next we determine the {\em second} random choice matrix $M_{\{\eta^{(2)} \}} $
in the same manner using 
$\mid \psi^{(1)} \rangle = \sum \mid  i \rangle c^{(1)}_i$ 
instead of $\mid  \psi^{(0)} \rangle $, namely using the probability
functions 
\begin{eqnarray}
P_i^{(2)}(\eta) \equiv P(\eta ; a^{(2)}_i) \ ,
\ \ \ \ \ \ a^{(2)}_i \equiv 
\left\{ \begin{array}{ll} 
{\rm max} \left( 1, \epsilon/|c^{(1)}_i| \right)  & (c^{(1)}_i \neq 0 ) \\ 
\infty & (c^{(1)}_i = 0) \end{array} \right. \ .     
\label{pi2}
\end{eqnarray}
Then we proceed to calculate 
$\hat{H} M_{\{\eta ^{(2)} \}} \mid \psi^{(1)}\rangle $, $C^{(2)}$ and  
$\mid \psi^{(2)}\rangle$.
Repeating this procedure we obtain the $m$-th normalized state 
$\mid \psi^{(m)}\rangle$ from the preceding state $\mid
\psi^{(m-1)}\rangle$,
\begin{eqnarray}
\mid \psi^{(m)} \rangle &\equiv&
\hat{H} M_{\{\eta ^{(m)} \}} \mid \psi^{(m-1)}\rangle / C^{(m)} \ , \\   
\left[ C^{(m)} \right] ^2 &=& \langle \psi^{(m-1)} \mid  M_{\{\eta ^{(m)} \}}
 \hat{H}^2 M_{\{\eta ^{(m)} \}} \mid \psi^{(m-1)} \rangle \ .
\end{eqnarray}
The on-off probability functions to generate $M_{\{\eta^{(m+1)} \}}$ 
are defined using the coefficients $c_i^{(m)}$ in $\mid \psi^{(m)} \rangle$,
\begin{eqnarray}
\mid \psi^{(m)} \rangle &=& \sum \mid  i \rangle c^{(m)}_i \ ,
\label{exppm}
\end{eqnarray}
as 
\begin{eqnarray}
P_i^{(m+1)} \left( \eta \right) \equiv P(\eta ; a^{(m+1)}_i) \ ,
\ \ \ \ \ \ a^{(m+1)}_i \equiv 
\left\{ \begin{array}{ll} 
{\rm max} \left( 1, \epsilon/|c^{(m)}_i| \right)  & (c^{(m)}_i \neq 0 ) \\ 
\infty & (c^{(m)}_i = 0) \end{array} \right. \ .  
\label{pimp1}
\end{eqnarray}
Thus in the RSSS method 
we obtain for $E(L)_{\{ \eta \}} $ defined by (\ref{eldeff}), 
\begin{eqnarray}
E(L)_{\{ \eta \}} = \langle \psi^{(0)} \mid   \psi^{(L)} \rangle
C^{(L)}C^{(L-1)} \cdots C^{(1)} \ ,
\end{eqnarray}
for each set of the random values of  
$\{ \eta \} = \{\{ \eta^{(L)} \}, \cdots, \{ \eta^{(1)} \} \}$.

Note that we should be careful here to think of the statistical
averages concerning $\{\eta^{(m)}\}$ because the probability functions
$P_i^{(m)} \left( \eta \right) $'s
depend on the precedingly generated sets of random variables  
$\{ \eta^{(1)} \}$, $\{ \eta^{(2)} \}$, $\cdots$, $\{ \eta^{(m-1)} \}$.
Yet, we see that  
$\langle \! \langle \eta_i^{(m)} \rangle \! \rangle =1 $ holds even in
this case for a
{\em fixed} set of $\{ \{\eta^{(m-1)}\}, \cdots ,\{\eta^{(1)}\} \} $.
We therefore obtain the relation (\ref{eleel})
if we do recursive samplings which ensure  
\begin{eqnarray}
\langle \! \langle \langle \! \langle  \cdots 
\langle \! \langle \langle \! \langle \eta_i^{(L)} \rangle \! \rangle_
{{\rm fixed}\{ \{\eta^{(L-1)}\}, \cdots ,\{\eta^{(1)}\} \}} \ 
\eta_j^{(L-1)}\rangle \! \rangle_
{{\rm fixed}\{ \{\eta^{(L-2)}\}, \cdots ,\{\eta^{(1)}\} \}} \ \cdots
\eta_k^{(2)}\rangle \! \rangle_{{\rm fixed} \{\eta^{(1)}\}} \
\eta_l^{(1)}\rangle \! \rangle =1 \ .
\label{recave}
\end{eqnarray}
We expect that variances of 
$\langle \! \langle E_{{\{\eta \}}}(L) \rangle \! \rangle$ in the RSSS
method become smaller for large $L$ than those in the SSS method 
because our definition of $a_i^{(m+1)}$ in (\ref{pimp1}) 
minimizes a quantity $S^{(m)} \equiv \epsilon^2 \langle \! \langle N^{(m)} 
\rangle \! \rangle + \langle \! \langle \left[g^{(m)}\right]^2
\rangle \! \rangle $. Here $g^{(m)}$ measures 
the difference of the truncated state $M_{\{ \eta ^{(m+1)} \}} 
\mid \psi^{(m)} \rangle$ from the original one $\mid \psi^{(m)} \rangle$,   
\begin{eqnarray}
\mid \chi^{(m)} \rangle \ g^{(m)} \equiv M_{\{ \eta ^{(m+1)} \}} 
\mid \psi^{(m)} \rangle \ - \mid \psi^{(m)} \rangle \ , \ \ \ \ \
\langle \chi^{(m)} \mid \chi^{(m)} \rangle = 1 \ ,
\label{chim} 
\end{eqnarray} 
and $N^{(m)} $ denotes the number of non-zero coefficients in (\ref{exppm}).
The reason why $S^{(m)}$ is minimized by $a_i^{(m+1)}$ in (\ref{pimp1}) 
will be given in the Appendix. 

\section{Numerical Examinations}

Here we numerically examine our method. We measure 
\begin{eqnarray}
 \langle \! \langle E_{{\{\eta \}}}(L) \rangle \! \rangle_{\rm smpl} &\equiv&
\frac{1}{n_{\rm smpl}} \sum_{k=1}^{n_{\rm smpl}} E_{{\{\eta \}_k}}(L) \ ,
\label{wem4}
\end{eqnarray}
in order to evaluate $E(L)$, where $\{\eta \}_k$
denotes the $k$-th set of random values of  
$\{\{\eta^{(L)}\}, \cdots, \{\eta^{(1)}\} \}$. 
In the SSS method, replacing $\langle \! \langle E_{{\{\eta \}}}(L) \rangle \!
\rangle$ by (\ref{wem4}) is theoretically allowed because 
$\{\eta^{(m)}\}$ is independent of $\{\eta^{(m')}\}$ if $m \neq m'$. 
In the RSSS method, on the other hand, things are more
complicated as was pointed out in the previous section. 
Since the recursive sampling to ensure (\ref{recave}) is 
difficult to carry out in numerical work 
because of the bursting sample numbers for 
large values of $L$, we adopt a sequential generation of only one set of 
$\{\eta\}_k =  \{ \{\eta^{(L)}\}, \cdots ,\{\eta^{(1)}\} \}_k$ for each
sampling in (\ref{wem4}). 
This replacement is analytically justified in the sense that an iterated
integral is replaceable by a multiple integral in most cases. 
We will see our numerical check 
confirms that we can obtain reliable values of $E(L)$ by (\ref{wem4}) 
without recursive samplings.

After the same manner as the SSS method, we evaluate the error of 
$ \langle \! \langle E_{{\{\eta \}}}(L) \rangle \! \rangle _{\rm smpl}$
by
\begin{eqnarray}
Er(L) &\equiv& 2 
\sqrt{\frac{ \rho_{{\{\eta \}}}^2(L)}{n_{\rm smpl}}} \ , 
\end{eqnarray}
where  
\begin{eqnarray}
 \rho_{{\{\eta \}}}^2(L) & \equiv &
 \langle \! \langle \left[ E_{\{\eta \}}(L)\right]^2 
\rangle \! \rangle _{\rm smpl} -
\left[ \langle \! \langle E_{\{\eta \}}(L) 
\rangle \! \rangle _{\rm smpl}\right]^2 \ .
\label{wem5}
\end{eqnarray} 

As a concrete example we adopt 
the anti-ferromagnetic Heisenberg quantum spin system on the triangular
lattice. The Hamiltonian of the model is 
\begin{eqnarray}
\hat H = \frac{J}{4} \sum_{(i,j)} \mbox{\boldmath $\sigma$}_i \cdot 
\mbox{\boldmath $\sigma$}_j \ ,
\end{eqnarray}
where \mbox{\boldmath $\sigma$}$_i$ is the Pauli matrix on the site $i$
and $(i,j)$ runs over all nearest neighbor pairs on the lattice
with the periodic boundary conditions for both directions\cite{bernu,foot2}. 
The coupling $J$ is fixed to be 1 throughout this paper.

First we present in Tables I and II our results for a small lattice with
21 sites. Since we can obtain an exact ground state 
$\mid \psi_{\rm E} \rangle$ by the Lanczos method for this system, we put 
$\mid \psi^{(0)} \rangle = \mid \psi_{\rm E} \rangle$ here with  
$N^{(0)}=352,716$. 
Table I shows results on 
$\langle \! \langle E_{{\rm E}\{\eta \}}(L) \rangle \! \rangle _{\rm smpl}$ 
up to $L=15$ together with the exact values $E_{\rm E}(L)$, 
where we add a suffix E in order to indicate that an exact eigenstate 
is used as $\mid \psi^{(0)} \rangle$. The value of the parameter
$\epsilon$ is 0.1. 
In Table II we present 
$\langle \! \langle  N^{\rm E}_{\rm b}(L) \rangle \! \rangle _{\rm smpl}$
and
$\langle \! \langle  N^{\rm E}_{\rm a}(L) \rangle \! \rangle _{\rm smpl}$, 
which denote the numbers of non-zero coefficients
before and after operating the random choice matrix
$M_{\{ \eta ^{(L)}\}}$ to the state
$\hat H M_{\{ \eta ^{(L-1)}\}} \cdots \hat H
M_{\{ \eta ^{(1)}\}} \mid  \psi_{\rm E} \rangle $.
For comparison, these Tables also contain the 
results obtained by the SSS method. 
The value of the parameter $\epsilon$ in the SSS method 
is adjusted to be 0.06 so that 
$\langle \! \langle  N^{\rm E}_{\rm b}(L) \rangle \! \rangle _{\rm smpl}$
nearly equals to that in the RSSS method. In fact, Table II shows that
$\langle \! \langle  N^{\rm E}_{\rm b}(L) \rangle \! \rangle _{\rm smpl}
\sim O(5 \times 10^4)$ for $L \geq 6$ in the SSS method, 
which is comparable with 
$\langle \! \langle  N^{\rm E}_{\rm b}(L) \rangle \! \rangle _{\rm smpl}
\sim O(6 \times 10^4)$ for $L \geq 3$ in the RSSS method. 
We see that the RSSS results in Table I fairly reproduce the exact values and 
that, as is expected, errors are less than those in the SSS results.
Although $Er(L)$ is smaller in the SSS method
when $L \leq 3$ due to the smaller value of $\epsilon$, 
it becomes smaller in the RSSS method for $L \geq 4$. The difference 
between errors in both methods grows as $L$ increases.  
The RSSS method therefore surely enables us to obtain meaningful 
results for larger values of $L$ compared to the SSS method.

Now we turn to the larger size system of 36 sites.
For our purpose to evaluate the energy eigenvalue
based on the power method, we first have to find an approximate state
for this system.
Here we use the restructuring techniques\cite{mmptp}
to form fundamental basis states where three spins of each of 
non-overlapping triangles are diagonalized.
Since neither simple way employed in the previous
work\cite{mune,mune2} could not give us a good approximate state, 
we combine them.  
Referring to the discussions in ref.~13
we start from a trial state and repeat the following procedure 
until the change of the obtained eigenvalue becomes negligibly small.

\begin{enumerate}
\item Operate the Hamiltonian to the trial state repeatedly 
as long as the number of the basis states is
acceptable to our computer memory resources. Then a truncated
      vector space is formed. 
\item Calculate the lowest energy eigenvalue in the truncated vector space
      by means of the Lanczos method.
\item Drop small coefficients of the obtained eigenstate. Use it as 
      the next trial state.
\end{enumerate}

Our best approximate state 
$\mid \psi_{\rm A} \rangle$ obtained from the above procedure is composed of 
12,281,253 non-zero components. The expectation value of $\hat H$ is 
$\langle \psi_{\rm A} \mid \hat H \mid \psi_{\rm A} \rangle = 
-18.418$\cite{foot3}.
Using this $\mid \psi_{\rm A} \rangle$ we calculate the expectation values. 
Let $Q_{\rm A}(L) \equiv 
\langle \psi_{\rm A}\mid \hat Q^L  \mid \psi_{\rm A}\rangle $, 
where  $\hat Q \equiv 5\hat I -\hat{H}$ with the identity operator $\hat I$.
Since our method is based on the power method, we use  
$\hat Q ^L$ instead of $\hat H^L$ in order to suppress excited states.
Results on $ \langle \! \langle Q_{{\rm A} \{\eta \}}(L) 
\rangle \! \rangle_{\rm smpl}$ as well as 
$\langle \! \langle  N^{\rm A}_{\rm b}(L) \rangle \! \rangle _{\rm
smpl}$ and $\langle \! \langle  N^{\rm A}_{\rm a}(L) \rangle \! \rangle _{\rm
smpl}$ with $\epsilon = 0.016 $ and $n_{\rm smpl} = 10^3$ 
are presented in Table III up to $L = 12$.
Here the maximum value of $L$ is determined under a condition that
the square root of the variance, $\sqrt{\rho^2_{\{ \eta \}} (L)}$, 
is less than the expectation value.
The maximum program size is about one Gbytes and 
the CPU time is about one hour for one sampling by a Pentium 4 machine.
The results in Table III show that the RSSS method is effective, not only for 
$\mid \psi _{\rm E} \rangle$ of the 21-site system but also for 
$\mid \psi _{\rm A} \rangle$ of the 36-site
one, to suppress a rapid increase of the variance for large $L$. For
example, the variance $\rho_{{\{\eta \}}}^2(5)$ in the SSS method
is $9.3 \times 10^{13}$, which is about 600 times as large as the
RSSS result $\rho_{{\{\eta \}}}^2(5)=1.6 \times 10^{11}$.
We also see in Table III that the number of non-zero coefficients
drastically reduces by the random choice matrix,  
$\langle \! \langle  N^{\rm A}_{\rm a}(L) \rangle \! \rangle _{\rm smpl}
\ll \langle \! \langle  N^{\rm A}_{\rm b}(L) \rangle \! \rangle _{\rm
smpl}$ for each $L$, and it increases very slowly as $L$ grows.  
In Figure~1 we plot the ratio $\langle \! \langle Q_{{\rm A} {\{\eta \}}}(L) 
\rangle \! \rangle_{\rm smpl} / 
\langle \! \langle Q_{{\rm A} {\{\eta \}}}(L-1) \rangle \! \rangle_{\rm
smpl}$ $(\langle \! \langle Q_{{\rm A} {\{\eta \}}}(0) 
\rangle \! \rangle_{\rm smpl} \equiv 1)$, 
which should give the exact value $5-E$ in the large $L$ limit.
We observe that the ratio increases for the data up to $L =10$. 
This increase suggests that the power method works.
For $L=11$ and 12 the errors are too large to
see this tendency. We therefore exclude   
$\langle \! \langle Q_{{\rm A} \{\eta \}}(11) \rangle \! \rangle_{\rm
smpl}$ and 
$\langle \! \langle Q_{{\rm A} \{\eta \}}(12) \rangle \! \rangle_{\rm
smpl}$ in the following analysis.

In order to estimate the energy eigenvalue from these expectation values
we use the same fitting form as was introduced in the previous 
work\cite{mune},   
\begin{eqnarray} 
\langle \psi_{\rm A}| \hat Q ^L | \psi_{\rm A}\rangle 
= Q^L \left(q_0 + \frac{q_1}{L +\alpha +1} \right) \equiv 
F(L,Q,q_0,q_1,\alpha) \ ,
\label{fitf}
\end{eqnarray}  
$q_0$, $q_1$, $\alpha$ being free parameters which are to be 
determined together with $Q$ by the fit. 
Changing these four parameters under two constraints $q_0 + q_1/(\alpha+1)
 = \langle \psi_{\rm A} \mid \psi_{\rm A}\rangle = 1$
and $Q \{q_0 + q_1/(\alpha+2)\} = 
\langle \psi_{\rm A}| \hat Q | \psi_{\rm A}\rangle = -18.418$, 
we look for the minimum of the difference $D$,
\begin{eqnarray}
D \equiv \sum_{L=2}^{L_{\rm max}}
\left[  1- \frac{ \ \langle \! \langle Q_{{\rm A}{\{\eta \}}}(L) 
\rangle \! \rangle _{\rm smpl}}{F(L,Q,q_0,q_1,\alpha)} \right]^2,
\label{diffd}
\end{eqnarray} 
with $L_{\rm max}=10$. 
We accept the values of the parameters 
if $D$ is less than the sum of the relative errors 
\begin{eqnarray*}
\sum_{L=2}^{L_{\rm max}} \left[
\frac{Er(L)}{ \ \langle \! \langle Q_{{\rm A}{\{\eta \}}}(L) \rangle \!
\rangle _{\rm smpl}}\right] ^2 \ = 2.2 \times 10^{-3} \ . 
\end{eqnarray*} 
The result of this fit is $E_{\rm fit} = 5 - Q_{\rm fit} = -20.50 
\pm \begin{array}{l} 
{\mbox{\small 1.37}} \\ {\mbox{\small 0.12}} \end{array}$.  
The function $F(L,Q_{\rm fit},q_{0 {\rm fit}},
q_{1 {\rm fit}},\alpha_{\rm fit})$ with fitted parameters to minimize
$D$ is plotted with a solid line in Fig.~1.

\section{Further Calculations}

In this section we add several results for the 36-site system.
In the previous section we obtained, starting from an approximate state
$\mid \psi _{\rm A} \rangle$, the ground state energy of the 36-site
system which is in agreement with the known exact value. Statistical
errors, however, are not satisfactorily small. One way to decrease the
errors is to increase the number of samples, but it is quite
time-consuming. In this section we employ another way, which is to
improve the trial state by requesting the translational invariance.
For this purpose we introduce a wave vector $\mbox{\boldmath  ${\rm k}$}$. 
Roughly every $N_{\rm s}$ basis states are linearly combined 
to form one new basis state for a definite value of 
$\mbox{\boldmath  ${\rm k}$}$, where $N_{\rm s}$ denotes the number of 
sites of the system. Requesting $\mbox{\boldmath  ${\rm k}$}= 
\mbox{\boldmath  $0$}$ we construct an improved trial state $\mid \psi
_{\rm T} \rangle$, whose number of new basis states amounts $13,911,394$
and $\langle \psi_{\rm T} \mid \hat H \mid \psi_{\rm T} \rangle =
-19.710$. 
Here we abandon the restructuring techniques 
in order to avoid too much complexity.
Results up to $L=6$ from $140$ samples with $\epsilon = 0.02$
are shown in Table IV and Fig.~1. 
Using the same fitting form (\ref{fitf}) we obtain
$E_{\rm fit} = -20.04 \pm \begin{array}{l} 
{\mbox{\small 0.10}} \\ {\mbox{\small 0.06}} \end{array}$, whose
statistical errors are less than $0.5 \%$.
The number of non-zero coefficients before (after) operating the
$6$-th random choice matrix is
$\langle \! \langle  N^{\rm T}_{\rm b}(6) \rangle \! \rangle _{\rm smpl}
\sim 3.0 \times 10^7$ $(\langle \! \langle  N^{\rm T}_{\rm a}(6) \rangle
\! \rangle _{\rm smpl} \sim 7.9 \times 10^5)$.

The lowest energy eigenvalues for excited states with fixed values of
$S_z$, the $z$-component of the total spin, are also calculable in similar 
manners. Our results for the $S_z=1$ homogeneous 
($\mbox{\boldmath  ${\rm k}$}= \mbox{\boldmath  $0$}$) state are
presented in Table V. Here
$\epsilon = 0.02$ and $n_{\rm smpl} = 88$ for $L \leq 5$, while 
$\epsilon = 0.03$ and $n_{\rm smpl} = 180$ for $L \geq 6$.
We start from a trial state 
$\mid \psi_{\rm T} (S_z=1) \rangle$, for which 
$\langle\psi_{\rm T} (S_z=1) \mid \hat H  \mid \psi_{\rm T} (S_z=1) 
\rangle = -18.813$. Up to $L=5$ we, as usual, calculate inner products of 
$\langle\psi_{\rm T} (S_z=1) \mid$ and 
$\hat Q M_{\{\eta^{(L)}\}} \cdots  \hat Q M_{\{\eta^{(1)}\}}  
\mid \psi_{\rm T} (S_z=1) \rangle$. For $L \geq 6$, on the other hand, 
we measure inner products between {\em independently
generated} $\langle\psi_{\rm T} (S_z=1) \mid M_{\{\eta^{'(1)}\}}\hat Q
\cdots M_{\{\eta^{'(L-5)}\}} \hat Q $ and 
$\hat Q M_{\{\eta^{(5)}\}} \cdots \hat Q  M_{\{\eta^{(1)}\}}  
\mid \psi_{\rm T} (S_z=1) \rangle$ in order to save memory resources. 
This measurement is allowed because 
$\langle \! \langle \ \langle \psi \mid 
M_{\{\eta^{'(1)}\}} \hat Q \cdots M_{\{\eta^{'(L-5)}\}} \hat Q 
\cdot \hat Q M_{\{\eta^{(5)}\}} \cdots \hat Q M_{\{\eta^{(1)}\}} 
\mid \psi \rangle \ \rangle \! \rangle = 
\langle \! \langle \ \langle \psi \mid \hat Q^L \mid \psi \rangle
\ \rangle \! \rangle$ holds\cite{footj}.
Based on the assumption (\ref{fitf}), we obtain
$E_{\rm fit} = -19.21 \pm \begin{array}{l} 
{\mbox{\small 0.27}} \\ {\mbox{\small 0.10}} \end{array}$
from data in Table V.

Finally, in order to demonstrate that we can extract physical properties
of the ground state with this method, 
we report results on the static structure factor 
$F(\mbox{\boldmath  ${\rm k}$})$ defined by 
\begin{eqnarray}  
F(\mbox{\boldmath ${\rm k}$}) \equiv \langle \psi_{\rm E} \ | \
\mbox{\boldmath ${\rm S}$}_{-\mbox{\boldmath ${\rm k}$}} \cdot
\mbox{\boldmath ${\rm S}$}_{ \mbox{\boldmath ${\rm k}$}} \ | \ \psi_{\rm E}
\rangle \ , \ \ \ \ \ \mbox{\boldmath ${\rm S}$}_{\mbox{\boldmath ${\rm k}$}}
\equiv \frac{1}{2\sqrt{N_{\rm s}}} \sum_j 
\mbox{\boldmath $\sigma$}_j e^{-i\mbox{\boldmath ${\rm k}$}  
\mbox{\boldmath ${\rm r}$}_j} \ ,
\label{fk}
\end{eqnarray}
where $j$ runs over all site of the lattice and 
$\mbox{\boldmath ${\rm r}$}_j$ denotes the position vector for site
$j$. In numerical study we define
\begin{eqnarray}
F_m(\mbox{\boldmath ${\rm k}$}) \equiv \frac{\langle \psi_{\rm T} \ | \
\hat Q ^m \ \mbox{\boldmath ${\rm S}$}_{-\mbox{\boldmath ${\rm k}$}} \cdot
\mbox{\boldmath ${\rm S}$}_{\mbox{\boldmath ${\rm k}$}} \ \hat Q ^m \ | \ 
\psi_{\rm T} \rangle } {\langle \psi_{\rm T} \ | \
\hat Q ^{2m} \ | \ \psi_{\rm T} \rangle } \ ,
\label{fmk}    
\end{eqnarray}
which gives $F(\mbox{\boldmath ${\rm k}$})$ in the large $m$ limit,
\begin{eqnarray}
\lim _{m \rightarrow \infty}F_m(\mbox{\boldmath ${\rm k}$}) =
F(\mbox{\boldmath ${\rm k}$}) \ .
\label{fmklim}
\end{eqnarray}
Then we evaluate $F_m(\mbox{\boldmath ${\rm k}$})$ from measurements of  
\begin{eqnarray}
\frac {\langle \! \langle \ \langle \psi_{\rm T} \ | \
\hat Q M_{\{\eta^{(2m-2)}\}} 
\cdots \hat Q M_{\{\eta^{(m)}\}} \ 
\mbox{\boldmath ${\rm S}$}_{-\mbox{\boldmath ${\rm k}$}} \cdot
\mbox{\boldmath ${\rm S}$}_{ \mbox{\boldmath ${\rm k}$}} 
\ \hat Q M_{\{\eta^{(m-1)}\}} 
\cdots \hat Q M_{\{\eta^{(1)}\}} \ | \ 
\psi_{\rm T} \rangle \ \rangle \! \rangle _{\rm smpl} }
{\langle \! \langle  \ \langle \psi_{\rm T} \ | \ 
\hat Q M_{\{\eta^{(2m-2)}\}} \cdots \hat Q M_{\{\eta^{(m)}\}} 
\hat Q M_{\{\eta^{(m-1)}\}} \cdots \hat Q M_{\{\eta^{(1)}\}}
\ | \ \psi_{\rm T} \rangle \ \rangle \! \rangle _ {\rm smpl}} \ .
\label{fmkmes}
\end{eqnarray}
Our results for $\epsilon = 0.03$, $n_{\rm smpl}=8$ and $m=6$ are
shown in Fig. 2. Here we do not need many samples because we allow
maximally one per cent of the statistical errors for $F_m(\mbox{\boldmath ${\rm
k}$})$. They are in good agreement with the data from ref.~7 except for 
those with $k/k_0=1$. When $k=k_0$ we can obtain
a better value from the squared sublattice magnetization ${\cal M} ^2$,
using the relation\cite{footi} 
$2 N_{\rm s}F(\mbox{\boldmath ${\rm k}$}_0) = 9 {\cal M}^2$. 
The reason for this better value is that 
in measurement of ${\cal M}^2$ one does not suffer from heavy cancellations. 
We find this value is much closer to the datum given in ref.~7.  

\section{Summary and Discussions}

In this paper we developed the recursive stochastic state selection
(RSSS) method, which is derived from 
the stochastic state selection (SSS) method\cite{mune}.
As the word ``recursive'' indicates,
on-off probability functions to generate the random choice matrix 
are recurrently determined so that they reflect the 
newest intermediate state at each step of the RSSS procedure.
The merit of this modification is that variances for expectation values with
high powers of the Hamiltonian grow slowly compared with those in the 
SSS method. 

As a result of less variances we can now evaluate the ground state energy for
the triangular lattice Heisenberg spin-$1/2$ system
by means of the stochastic selections.
On a relatively small 21-site lattice we examined the RSSS method using 
the exact eigenstate. Results there
show that the variance $\rho_{\{ \eta \}}^2(10)$  
was reduced to $ \sim 1/2000$ of the one in the SSS method.
On a larger 36-site lattice we calculated the expectation values of 
$(5\hat I - \hat H) ^{L}$  
from two approximate states $\mid \psi _{\rm A} \rangle$ and 
$\mid \psi _{\rm T} \rangle$, where the latter realizes the wave vector 
$\mbox{\boldmath  ${\rm k}$}= \mbox{\boldmath  $0$}$. 
We again emphasize that the numerical study of such high powers would
not become possible without introducing the RSSS method. 
From obtained data with $\mid \psi _{\rm T} \rangle$ 
we estimated the energy of the ground state $E$ 
based on the extrapolation assumption.
The result indicates $-20.10 \leq E \leq -19.94$, which is slightly
higher than the known exact value $-20.17344$.
On the 36-site triangular lattice we also calculated expectation values
on the lowest energy eigenvalue with the $S_z =1$ state.
Further, we calculated the static structure factor, which 
is a typical measurement to obtain physical properties of the ground state. 
Our results fully ensure that not only energy eigenvalues but also 
the static structure factor is calculable in our method.  

Let us finally comment on whether it is possible to study larger
systems, the 48-site system for instance, by this method. In a simple 
estimation by a guess that the cost is proportional to the system size,
we would need $2^{48}/2^{36} \sim 4 \times 10^3$ times of computer
resources compared to the the CPU time and the memory we used in our 
present study for the 36-site system. This factor is too large for us 
to easily overcome. We, however, show in this study that a
better trial state with a fixed $\mbox{\boldmath  ${\rm k}$}$ 
can much reduce the cost of calculations. We hope, therefore, we can 
make a numerical study of larger triangular spin systems in the 
near future.           
 
\appendix
\section{On-Off Probability Functions}

In this Appendix we consider a random
choice matrix $M_{ \{ \eta \}} \equiv {\rm diag.} \{ \eta_1, \eta_2,
\cdots, \eta_{N_{\rm V}}\}$, where each random variable $\eta_i$ is
generated by an on-off probability function defined with a parameter 
$\xi_i$ $( \geq 1)$,
\begin{eqnarray}
 P_i(\eta)  \equiv P(\eta; \xi_i) = \frac{1}{\xi_i} \delta \left( \eta - \xi_i \right) + \left( 1 - \frac{1}{\xi_i} \right) \delta \left( \eta \right) \ . 
\label{pixii}
\end{eqnarray}
Our purpose is to discuss what values of 
$\{ \xi \} \equiv \{ \xi_1, \xi_2, \cdots,\xi_{N_{\rm V}}\}$ are
favorable in order to decrease the variances of the expectation values.
We will see that the definition of the on-off probability functions in
the RSSS method is ``best'' in the sense shown below.

First, let $N_i$ denote the number of non-zero $\eta_i$, which is therefore 
1 (0) when $\eta_i \neq 0$ ($\eta_i = 0$). 
Following the discussion in ref.~1, we
obtain for the statistical average of $N \equiv \sum N_i$
in the above $M_{ \{ \eta \}}$, 
\begin{eqnarray}
\langle \! \langle  N  \rangle \! \rangle = \sum_{i=1}^{N_{\rm V}}  
\frac{1}{\xi_i} \ ,
\label{nave}
\end{eqnarray}
because
\begin{eqnarray}
\langle \! \langle  N_i  \rangle \! \rangle &\equiv& \int_0^{\infty}  
N_i  P_{N_i}( N_i) d  N_i = \frac{1}{\xi_i} 
 \ ,  \\
P_{N_i}( N_i) &\equiv&
 \frac{1}{\xi_i}\delta( N_i -1)+(1-\frac{1}{\xi_i})\delta( N_i) \ . 
\end{eqnarray}
Next, multiplying $M_{ \{ \eta \}}$ to a normalized state vector 
$\mid \Psi \rangle$,
\begin{eqnarray}
\mid \Psi \rangle = \sum_{i=1}^{N_{\rm V}} \mid i \rangle c_i  \ ,
\label{dppsi}
\end{eqnarray} 
we obtain 
\begin{eqnarray}
 M_{\{ \eta \}}\mid \Psi \rangle &=& 
\sum_i \mid  i \rangle c_i \eta_i \ , 
\label{expmp}
\\
\langle \Psi \mid \left[  M_{\{ \eta \}} \right]^2 
\mid \Psi \rangle &=& \sum_i c_i^2  \eta_i^2   
= 1 + \sum_i c_i^2  \left( \eta_i ^2 - 1 \right)  \ ,
\label{pmmp1}
\end{eqnarray}
where the normalization condition $\langle \Psi \mid \Psi\rangle = \sum
c_i^2 = 1$ is used.
Let us then introduce a state vector $\mid \chi \rangle g$
which represents the difference between $M_{ \{ \eta \}} \mid \Psi \rangle$
and $\mid \Psi \rangle$,
\begin{eqnarray}
\mid \chi \rangle g \equiv M_{ \{ \eta \}} \mid \Psi \rangle - 
\mid \Psi \rangle = \sum_i \mid i \rangle c_i \left( 
\eta_i -1 \right) \ ,
\label{chidef}
\end{eqnarray}
with a normalized $\mid \chi \rangle$.
It should be emphasized here that 
$ \langle \! \langle \ \langle \Phi \mid \chi \rangle g \   
\rangle \! \rangle = 0 $ 
holds for {\em any} state 
$\mid \Phi \rangle \equiv \sum \mid i \rangle b_i$ because 
$ \langle \! \langle \eta_i \rangle \! \rangle = 1$. 
From (\ref{chidef}) we see 
\begin{eqnarray}
 \langle \! \langle \ 
\langle \Psi \mid \left[ M_{ \{ \eta \} } \right]^2 \mid \Psi\rangle  \
\rangle \! \rangle
=  \langle \! \langle \
\langle \Psi \mid \Psi \rangle+ 2 g \langle \chi \mid \Psi \rangle
+ g ^2 \langle \chi \mid \chi \rangle  \ \rangle \! \rangle  
=  1+ \langle \! \langle g^2 \rangle \! \rangle.
\label{pmmp2} 
\end{eqnarray}
Using 
\begin{eqnarray}
\langle \! \langle  \eta^2_i  \rangle \! \rangle &\equiv& \int_0^{\infty}  
\eta^2_i  P_i (\eta_i) d  \eta_i = \xi_i 
\end{eqnarray}
we obtain from (\ref{pmmp1}) and (\ref{pmmp2}),
\begin{eqnarray}
\langle \! \langle g^2 \rangle \! \rangle = 
\langle \! \langle \sum_i c_i^2  \left(  \eta_i^2 - 1 \right)
\rangle \! \rangle 
= \sum_i c_i^2  \left(\langle \! \langle \eta_i^2
\rangle \! \rangle - 1 \right) = \sum_i c_i^2  \left( \xi_i -1\right) \ .
\label{gsq}
\end{eqnarray}
Is it possible to find any $\{ \xi \}$ which makes both of 
$\langle \! \langle N \rangle \! \rangle$
and $\langle \! \langle g^2 \rangle \! \rangle $ small?  
From (\ref{nave}) and (\ref{gsq}) we learn that we cannot lessen 
$\langle \! \langle g^2 \rangle \! \rangle $ without increasing $\langle
\! \langle N \rangle \! \rangle$.
Let us consider, then, to minimize a quantity 
\begin{eqnarray}
\epsilon^2 \langle \! \langle N
\rangle \! \rangle + \langle \! \langle g ^2 \rangle \! \rangle  
= \sum_{i=1}^{N_{\rm V}}  
\left\{ \frac{\epsilon^2}{\xi_i} +c_i^2  \left(\xi_i - 1 \right) \right\} 
\equiv S \left( \xi_1, \xi_2, \cdots, \xi_{N_{\rm V}} \right)  \ ,
\label{sdef}
\end{eqnarray} 
 with a positive constant
$\epsilon $ in ranges $\xi_i \geq 1$ for all $i$. 
Since $\partial S/\partial \xi_i = c_i^2 -\epsilon^2/\xi_i^2 $, we find
\begin{itemize}
\item $\partial S/\partial \xi_i = 0$ at
$\xi_i= \epsilon / |c_i|$ $( > 1)$ if $0 < |c_i| < \epsilon$.
\item  $\partial S/\partial \xi_i > 0 $ for 
$\xi_i \geq 1$ when $|c_i| > \epsilon$. In this case $\xi_i = 1$ gives
      the minimum of $\epsilon^2 / \xi_i + c_i^2(\xi_i -1)$. 
\item  $\partial S/\partial \xi_i < 0 $ if $c_i=0$. 
In this case $\epsilon^2 / \xi_i + c_i^2(\xi_i -1)
= \epsilon^2 / \xi_i \rightarrow 0$ when $\xi_i \rightarrow \infty$.
\end{itemize}
Therefore the minimum value of 
$S \left( \xi_1, \xi_2, \cdots, \xi_{N_{\rm V}} \right) $ is realized by  
\begin{eqnarray} 
\xi_i = \left\{ \begin{array}{ll} 
{\rm max} \left( 1, \epsilon/|c_i| \right)  & (c_i \neq 0 ) \\ 
\infty & (c_i = 0) \end{array} \right. \ .
\end{eqnarray} 

To summarize this Appendix, we find the ``best''
choice of $\{ \xi \}$ in (\ref{pixii}), by which the quantity 
$\epsilon^2 \langle \! \langle N
\rangle \! \rangle + \langle \! \langle g ^2 \rangle \! \rangle  $  
in (\ref{sdef}) is minimized.

\newpage
\begin{table}[ht]
\centering
\begin{tabular}{rrlrlrl} \hline
\multicolumn{1}{c}{ } &  \multicolumn{2}{c}{ }
&\multicolumn{2}{c}{RSSS}
  &\multicolumn{2}{c}{SSS}
\\ \cline{4-7}
\multicolumn{1}{c}{$L$}
& \multicolumn{2}{c}{$E_{\rm E}(L)$} & \multicolumn{4}{c}
{$\langle \! \langle E_{{\rm E}{\{\eta \}}}(L) \rangle \! \rangle _{\rm smpl}$}
 \\ \hline
 1  &  $-$0.117809 & $ \times  10^{2}$ &$-$ (0.117787 $\pm $ 0.000072) 
& $ \times 10^{2}$ & $-$(0.117796 $\pm $ 0.000050) &$ \times  10^{2}$ \\ \hline

 2  &  0.138790& $ \times  10^{3}$ & (0.13885 $\pm $ 0.00015) 
&$ \times  10^{3}$ & (0.138760 $\pm $ 0.000098) & $ \times  10^{3}$ \\ \hline

 3 & $-$ 0.163507& $ \times  10^{4}$& $-$(0.16362$\pm $ 0.00027) 
& $ \times  10^{4}$ & $-$(0.16343 $\pm $ 0.00020) &$ \times  10^{4}$  \\ \hline

 4  &  0.192626& $ \times  10^{5}$  & (0.19286 $\pm $ 0.00044) 
& $ \times  10^{5}$& (0.19244 $\pm $ 0.00067) &$ \times  10^{5}$ \\ \hline

 5  &   $-$0.226931& $ \times  10^{6}$& $-$(0.22710 $\pm $ 0.00070) 
& $ \times  10^{6}$ & $-$(0.2270$\pm $ 0.0030) &$ \times  10^{6}$  \\ \hline

 6  &  0.267346& $ \times  10^{7}$& (0.2671 $\pm $ 0.0011) 
& $ \times  10^{7}$ & (0.269$\pm $ 0.015) &$ \times  10^{7}$  \\ \hline

 7  &  $-$ 0.314958& $ \times  10^{8}$ & $-$(0.3152 $\pm $ 0.0018) 
& $ \times  10^{8}$& $-$(0.286$\pm $ 0.049) &$ \times  10^{8}$  \\ \hline

 8  &  0.371049& $ \times  10^{9}$ & (0.3708 $\pm $ 0.0030) 
& $ \times  10^{9}$& (0.24$\pm $ 0.17) &$ \times  10^{9}$  \\ \hline

 9  &  $-$ 0.437129& $ \times  10^{10}$& $-$(0.4357 $\pm $ 0.0050) 
& $ \times  10^{10}$ &$-$ (0.39$\pm $ 0.17) &$ \times  10^{10}$  \\ \hline

 10  &  0.514978& $ \times  10^{11}$& (0.5170 $\pm $ 0.0085) 
& $ \times  10^{11}$ & (0.30$\pm $ 0.41) &$ \times  10^{11}$  \\ \hline

 11  & $-$ 0.606691& $ \times  10^{12}$ &$-$ (0.607 $\pm $ 0.014) & 
$ \times  10^{12}$&\multicolumn{2}{c}{---}  \\ \hline

 12  &  0.714737& $ \times  10^{13}$ & (0.709 $\pm $ 0.025) 
& $ \times  10^{13}$ &\multicolumn{2}{c}{---} \\ \hline

 13  & $-$ 0.842025& $ \times  10^{14}$  & $-$(0.850 $\pm $ 0.042) 
& $ \times  10^{14}$ &\multicolumn{2}{c}{---} \\ \hline

 14  &  0.991982& $ \times  10^{15}$  & (1.005 $\pm $ 0.072) 
& $ \times  10^{15}$ & \multicolumn{2}{c}{---}  \\ \hline

 15  & $-$ 0.116865 & $ \times  10^{17}$ & $-$(0.122 $\pm $ 0.013) 
& $ \times  10^{17}$&\multicolumn{2}{c}{---}  \\ \hline

\end{tabular}
\caption{
Results on
$\langle \! \langle E_{{\rm E}{\{\eta \}}}(L) \rangle \! \rangle _{\rm smpl}$
$(L=1,2, \cdots, 15)$ obtained for the 21-site 
triangular lattice by the RSSS method with $\epsilon = 0.1$
and the SSS method with $\epsilon = 0.06$.
The number of samples is $10^4$.
Exact values of $E_{\rm E}(L)$ are also presented for comparison.}
\end{table}

\begin{table}[hb]
\centering
\begin{tabular}{rrrrr} \hline
 { }&\multicolumn{2}{c}{RSSS}
  &\multicolumn{2}{c}{SSS}   \\ \cline{2-5}
\multicolumn{1}{c}{$L$}
&$ \ \langle \! \langle  N^{\rm E}_{\rm b}(L) \rangle \! \rangle _{\rm smpl}$   & $ \ \langle \!
 \langle  N^{\rm E}_{\rm a}(L) \rangle \! \rangle _{\rm smpl} $  &
$ \ \langle \! \langle  N^{\rm E}_{\rm b}(L) \rangle \! \rangle _{\rm smpl}$ &
$ \ \langle \! \langle  N^{\rm E}_{\rm a}(L) \rangle \! \rangle _{\rm smpl} $ 
 \\ \hline
 1 & 352716.0   &2659.3  & 352716.0    & 4432.2    \\ \hline
 2 & 74254.9  & 2038.3 & 108023.1    & 2603.0    \\ \hline
 3 &60343.5   & 1923.6 & 70882.0   &  2197.7   \\ \hline
 4 & 57531.4  & 1920.4 & 60150.5    & 2048.7   \\ \hline
 5 & 57365.5  & 1946.4 & 56182.7    & 1984.9   \\ \hline
 6 & 57819.9  & 1975.7 & 54473.2    &  1955.8  \\ \hline
 7 & 58330.5  & 2000.4 & 53687.7    &   1941.7  \\ \hline
 8 &58775.9   & 2020.2 & 53304.5    &  1934.3   \\ \hline
 9 & 59672.5  & 2033.5 & 53102.0    & 1930.7    \\ \hline
10 & 60088.4  & 2043.6 &53003.4     & 1929.1   \\ \hline
11 & 60185.8  & 2050.4 & --- & --- \\ \hline
12 & 60143.4  & 2053.9 & --- & --- \\ \hline
13 & 60179.8  & 2057.8 & --- & --- \\ \hline
14 &  60624.2  & 2059.9 & --- & --- \\ \hline
15 & 60691.2  & 2061.4 & --- & --- \\ \hline
\end{tabular}
\caption{
Numbers of non-zero coefficients {\em before} 
($\langle \! \langle  N^{\rm E}_{\rm b}(L) \rangle \! \rangle _{\rm smpl}$)
and {\em after} 
($\langle \! \langle  N^{\rm E}_{\rm a}(L) \rangle \! \rangle _{\rm smpl}$) 
operating the random choice matrix $M_{\{ \eta ^{(L)}\}}$ to the state
$\hat H M_{\{ \eta ^{(L-1)}\}} \cdots \hat H
M_{\{ \eta ^{(1)}\}} \mid  \psi_E \rangle $
for the 21-site triangular lattice 
obtained from $10^4$ samples with $\epsilon = 0.1$ (RSSS)
or $\epsilon = 0.06$ (SSS).
}
\end{table}

\begin{table}[ht]
\centering
\begin{tabular}{rrlrr} \hline
\multicolumn{1}{c}{$L$}
& \multicolumn{2}{c}
{$\langle \! \langle Q_{{\rm A}{\{\eta \}}}(L) \rangle \! \rangle _{\rm smpl}$} & \multicolumn{1}{c}{$ \ \langle \! \langle  N^{\rm A}_{\rm b}(L) \rangle \! \rangle _{\rm smpl}$}   &\multicolumn{1}{c}{$ \ \langle \!
 \langle  N^{\rm A}_{\rm a}(L) \rangle \! \rangle _{\rm smpl} $  } 
 \\ \hline
 1  &  (0.234137 $\pm $ 0.000077) & $ \times 10^{2}$ &  12281253.0 &
 91378.8 \\ \hline
 2  &  (0.55286 $\pm $ 0.00045) & $ \times 10^{3}$ & 21738077.0 &
 69452.6 \\ \hline
 3  &  (0.13114 $\pm $ 0.00019) & $ \times 10^{5}$ & 19123511.3 & 
69429.2 \\ \hline
 4  &  (0.31260 $\pm $ 0.00071) & $ \times 10^{6}$ & 19821436.4 &
74755.3 \\ \hline
 5  &  (0.7469 $\pm $ 0.0025) & $ \times 10^{7}$ & 21556264.4 &
81933.1 \\ \hline
 6  &  (0.17941 $\pm $ 0.00093) & $ \times 10^{9}$ & 23674697.9 &
89782.3 \\ \hline
 7  &  (0.4296 $\pm $ 0.0035) & $ \times 10^{10}$ & 25924492.7 &
97767.5 \\ \hline
 8  &  (0.1028 $\pm $ 0.0014) & $ \times 10^{12}$ & 28180669.0 &
105616.8 \\ \hline
 9  &  (0.2507 $\pm $ 0.0055) & $ \times 10^{13}$ & 30379138.4 & 
113156.1 \\ \hline
10  &  (0.629 $\pm $ 0.024) & $ \times 10^{14}$ & 32478107.9 &
120316.8 \\ \hline
11  &  (0.153 $\pm $ 0.010) & $ \times 10^{16}$ & 34462103.9 & 
127006.7 \\ \hline
12  &  (0.355 $\pm $ 0.042) & $ \times 10^{17}$ & 36309064.0 &
133239.7 \\ \hline
\end{tabular}
\caption{
Results on
$\langle \! \langle Q_{{\rm A}{\{\eta \}}}(L) \rangle \! \rangle _{\rm smpl}$,
$ \ \langle \! \langle  N^{\rm A}_{\rm b}(L) \rangle \! \rangle _{\rm smpl}$ 
and 
$ \ \langle \! \langle  N^{\rm A}_{\rm a}(L) \rangle \! \rangle _{\rm smpl}$ 
$(L=1,2, \cdots, 12)$, obtained for the 36-site restructured
triangular lattice by the RSSS method with $\epsilon = 0.016$, where 
$Q_{\rm A}(L)
=\langle \psi_{\rm A} \mid (5 \hat I -\hat H)^L
\mid  \psi_{\rm A} \rangle $ with the identity operator $\hat I$.
The number of samples is $10^3$.}
\end{table}

\begin{table}[ht]
\centering
\begin{tabular}{rrlrr} \hline
\multicolumn{1}{c}{$L$}
& \multicolumn{2}{c}
{$\langle \! \langle Q_{{\rm T}{\{\eta \}}}(L) \rangle \! \rangle _{\rm smpl}$} & \multicolumn{1}{c}{$ \ \langle \! \langle  N^{\rm T}_{\rm b}(L) \rangle \! \rangle _{\rm smpl}$}   &\multicolumn{1}{c}{$ \ \langle \!
 \langle  N^{\rm T}_{\rm a}(L) \rangle \! \rangle _{\rm smpl} $  } 
 \\ \hline
 1  &  (0.247073 $\pm $ 0.000045) & $ \times 10^{2}$ & 13911394.0 &
 398861.8 \\ \hline
 2  &  (0.612899 $\pm $ 0.000193) & $ \times 10^{3}$ & 13513630.8 &
 437278.8 \\ \hline
 3  &  (0.152364 $\pm $ 0.000063) & $ \times 10^{5}$ & 16928438.7 & 
 518694.1 \\ \hline
 4  &  (0.37929 $\pm $ 0.00018) & $ \times 10^{6}$ & 21395583.2 &
 610462.3 \\ \hline
 5  &  (0.94546 $\pm $ 0.00057) & $ \times 10^{7}$ & 25728449.4 &
 701952.5 \\ \hline
 6  &  (0.23594 $\pm $ 0.00017) & $ \times 10^{9}$ & 30450059.3 &
 787238.1 \\ \hline
\end{tabular}
\caption{
Results on 
$\langle \! \langle Q_{{\rm T}{\{\eta \}}}(L) \rangle \! \rangle _{\rm smpl}$,
$ \ \langle \! \langle  N^{\rm T}_{\rm b}(L) \rangle \! \rangle _{\rm smpl}$ 
and 
$ \ \langle \! \langle  N^{\rm T}_{\rm a}(L) \rangle \! \rangle _{\rm smpl}$ 
$(L=1,2, \cdots, 12)$ obtained for the 36-site 
triangular lattice by the RSSS method with $\epsilon = 0.02$, where
a trial state $\mid \psi_{\rm T} \rangle$ is used instead of  
$\mid \psi_{\rm A} \rangle$.
The number of samples is $140$.}
\end{table}

\newpage
\begin{table}[ht]
\centering
\begin{tabular}{rrlrr} \hline
\multicolumn{1}{c}{$L$}
& \multicolumn{2}{c}
{$\langle \! \langle Q_{{\rm T}{\{\eta \}}}(L) \rangle \! \rangle _{\rm smpl}$} & \multicolumn{1}{c}{$ \ \langle \! \langle  N^{\rm T}_{\rm b}(L) \rangle \! \rangle _{\rm smpl}$}   &\multicolumn{1}{c}{$ \ \langle \!
 \langle  N^{\rm T}_{\rm a}(L) \rangle \! \rangle _{\rm smpl} $  } 
 \\ \hline
 1  &  (0.238085 $\pm $ 0.000067) & $ \times 10^{2}$ & 13957843.0 &
 557850.5 \\ \hline
 2  &  (0.569663 $\pm $ 0.000272) & $ \times 10^{3}$ &  17375161.1&
 553892.4 \\ \hline
 3  &  (0.136574 $\pm $ 0.000096) & $ \times 10^{5}$ & 20943597.7& 
 618879.1 \\ \hline
 4  &  (0.32779 $\pm $ 0.00026) & $ \times 10^{6}$ &  25154518.3 &
 699008.8 \\ \hline
 5  &  (0.78765 $\pm $ 0.00079) & $ \times 10^{7}$ & 29199658.1&
 779763.9 \\ \hline
 6  &  (0.18980 $\pm $ 0.00025) & $ \times 10^{9}$ & --- & ---  \\ \hline
 7  &  (0.45699 $\pm $ 0.00064) & $ \times 10^{10}$ & --- & ---  \\ \hline
 8  &  (0.11010 $\pm $ 0.00017) & $ \times 10^{12}$ & --- & ---  \\ \hline
 9  &  (0.26527 $\pm $ 0.00049) & $ \times 10^{13}$ & --- & ---  \\ \hline
10  &  (0.6401 $\pm $ 0.0015) & $ \times 10^{14}$ & --- & ---  \\ \hline
\end{tabular}
\caption{
Results on
$\langle \! \langle Q_{{\rm T}{\{\eta \}}}(L) \rangle \! \rangle _{\rm smpl}$
for $L=1,2, \cdots, 10$ and
$ \ \langle \! \langle  N^{\rm T}_{\rm b}(L) \rangle \! \rangle _{\rm smpl}$ 
and 
$ \ \langle \! \langle  N^{\rm T}_{\rm a}(L) \rangle \! \rangle _{\rm smpl}$ 
for $(L=1,2, \cdots, 5)$ in the $S_z=1$ case, obtained for the 36-site 
triangular lattice from 
a translational invariant trial state $\mid \psi_{\rm T} (S_z=1)\rangle$. 
Parameters are $\epsilon = 0.02$ and $n_{\rm smpl}=88$ up to $L=5$,
and $\epsilon = 0.03$ and $n_{\rm smpl}=180$ for $L \geq 6$.}
\end{table}

\begin{figure}[p]
\begin{center}
\scalebox{0.5}{\includegraphics{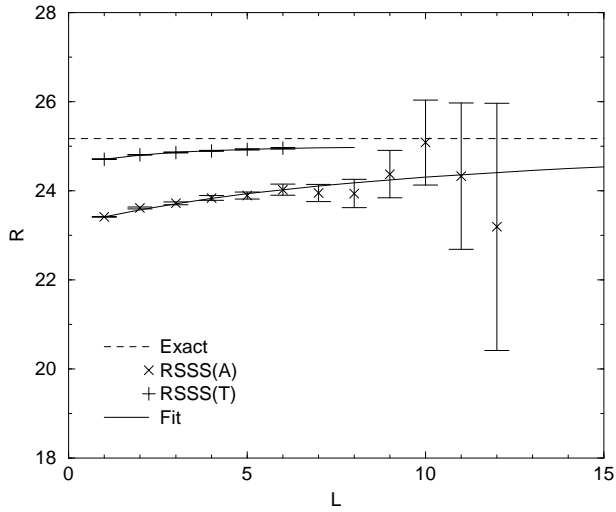}}
\caption{Ratios $R_{\rm A} \equiv \langle \! \langle Q_{{\rm A} 
{\{\eta \}}}(L) \rangle \! \rangle_{\rm smpl} / 
\langle \! \langle Q_{{\rm A} {\{\eta \}}}(L-1) \rangle \! \rangle_{\rm
smpl}$ (crosses) and $R_{\rm T} \equiv \langle \! \langle Q_{{\rm T} 
{\{\eta \}}}(L) \rangle \! \rangle_{\rm smpl} / 
\langle \! \langle Q_{{\rm T} {\{\eta \}}}(L-1) \rangle \! \rangle_{\rm
smpl}$ (pluses) for the Heisenberg spin model on the 36-site 
triangular lattice. Values of $R_{\rm A}$ are 
obtained from the results in Table III and 
$\langle \! \langle Q_{{\rm A} {\{\eta \}}}(0) 
\rangle \! \rangle_{\rm smpl} \equiv 1$. 
Values of $R_{\rm T}$ are obtained from the results in Table IV 
and $\langle \! \langle Q_{{\rm T} {\{\eta \}}}(0) 
\rangle \! \rangle_{\rm smpl} \equiv 1$. The dashed line indicates
the exact eigenvalue of $Q = 5 - E$\cite{leung, bernu,capri}. 
The solid lines present fitted values of (\ref{fitf}).} 
\end{center}
\end{figure}

\begin{figure}[p]
\begin{center}
\scalebox{0.5}{\includegraphics{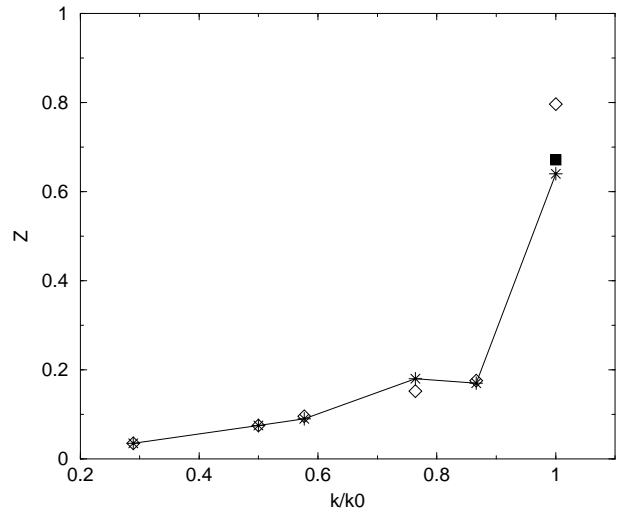}}
\caption{The static structure factor 
$Z \equiv 8F(\mbox{\boldmath  ${\rm k}$})/(N_{\rm s}+6)$ 
for the 36-site system versus $k/k_0$, where 
$\pm \mbox{\boldmath  ${\rm k}$}_0$ are the two wave vectors of 
the corners of the crystal Brillouin zone, 
$k \equiv |\mbox{\boldmath  ${\rm k}$}|$ and $N_{\rm s} (=36)$ is 
the number of sites.
Asterisks combined by lines to guide eyes are data from ref.~7. 
Our results obtained by 8 samples of $F_6(\mbox{\boldmath ${\rm k}$})$ 
with $\epsilon = 0.03$ are plotted by open diamonds. 
A filled square is the value calculated by  
$F(\mbox{\boldmath  ${\rm k}$}_0) = 4.5 {\cal M}^2/ N_{\rm s}$, where
${\cal M} ^2$ denotes the squared sublattice magnetization.}
\end{center}
\end{figure}


\begin{thebibliography}{99}
\bibitem{foot1}
In the SSS method we first defined $E_{\{\eta\}}(L)$ as
$$E_{\{\eta\}}(L) \equiv \langle \psi \mid M_{\{ \eta^{(L+1)}\}} \hat H
M_{\{ \eta ^{(L)}\}} \hat H M_{\{ \eta ^{(L-1)}\}} \cdots \hat H
M_{\{ \eta ^{(1)}\}} \mid  \psi \rangle \ ,$$
but in numerical calculations we often use 
$$E_{\{\eta\}}(L) \equiv \langle \psi \mid \hat H
M_{\{ \eta ^{(L)}\}} \hat H M_{\{ \eta ^{(L-1)}\}} \cdots \hat H
M_{\{ \eta ^{(1)}\}} \mid  \psi \rangle $$ 
in order to decrease statistical errors.
\bibitem{foot2}
For the 21-site lattice these directions are along vectors 
$\mbox{\boldmath $T$}_1 = 
4 \mbox{\boldmath $u$}_1+ \mbox{\boldmath $u$}_2 $ and 
$\mbox{\boldmath $T$}_2 = 
- \mbox{\boldmath $u$}_1+ 5 \mbox{\boldmath $u$}_2 $,
where $\mbox{\boldmath $u$}_1 = \left(1,0 \right)$ and 
$\mbox{\boldmath $u$}_2 = \left(1/2, \sqrt{3}/2 \right)$ are unit
vectors for the triangular lattice in the $x$-$y$ plain.  
For the 36-site lattice they are along  
$\mbox{\boldmath $T$}_1 = 6 \mbox{\boldmath $u$}_1 $ and 
$\mbox{\boldmath $T$}_2 = 6 \mbox{\boldmath $u$}_2 $.
\bibitem{foot3}
In the normal representation the state on each site of the lattice is 
represented by the $z$ component of the spin sitting on the site.
In this representation we obtain 
$\langle  \psi_{\rm A} ' \mid \hat H \mid \psi_{\rm A} ' \rangle = -17.6398$
for our best approximate state $\mid \psi_{\rm A} ' \rangle $.
\bibitem{footj}
For example, using (\ref{chim}) we see that 
\begin{eqnarray*}
\hat Q M_{\{\eta^{(1)}\}}  
\mid \psi \rangle &=& \hat Q \mid \psi \rangle + \hat Q \mid \chi \rangle g \ ,
\label{qmpsi} \\ 
\hat Q M_{\{\eta^{'(1)}\}}  
\mid \psi \rangle 
&=& \hat Q \mid \psi \rangle + \hat Q \mid \chi ' \rangle g' \ ,
\end{eqnarray*}
and then 
\begin{eqnarray*}
& & \langle \! \langle \
\langle \psi \mid 
M_{\{\eta^{'(1)}\}} \hat Q \cdot \hat Q M_{\{\eta^{(1)}\}} \mid \psi \rangle 
\ \rangle \! \rangle \nonumber \\ &=&
\langle \! \langle \ \langle \psi \mid \hat Q^2 \mid \psi \rangle
\ \rangle \! \rangle 
+ \langle \! \langle \ g' \langle \chi ' \mid \hat Q^2 \mid \psi \rangle 
\ \rangle \! \rangle 
+ \langle \! \langle \ \langle \psi \mid \hat Q^2 \mid \chi \rangle g
\ \rangle \! \rangle 
+ \langle \! \langle \ g' \langle \chi ' \mid \hat Q^2 \mid \chi \rangle g
\ \rangle \! \rangle \nonumber \\ 
&=& 
\langle \! \langle \ \langle \psi \mid \hat Q^2 \mid \psi \rangle
\ \rangle \! \rangle \ .  
\end{eqnarray*}
\bibitem{footi}
This is an identity for systems with the total spin 
$\mbox{\boldmath ${\rm S}$}=\mbox{\boldmath ${\rm 0}$}$\cite{bernu}.
\end{thebibliography}
\end{document}